\documentclass[twocolumn,showpacs,preprintnumbers,amsmath,amssymb]{revtex4}
\usepackage{graphicx} 
\usepackage{dcolumn}  
\usepackage{bm}       
\begin{document} 
\title{Correlated band structure of NiO, CoO and MnO by variational 
cluster approximation}
\author{R. Eder}
\affiliation{Forschungszentrum Karlsruhe, 
Institut f\"ur Festk\"orperphysik, 76021 Karlsruhe, Germany}
\date{\today}

\begin{abstract}
The variational cluster approximation proposed by Potthoff
is applied to the calculation of the single-particle spectral function of 
the transition metal oxides MnO, CoO and NiO.
Trial self-energies and the numerical value of the Luttinger-Ward functional
are obtained by exact diagonalization of a TMO$_6$-cluster. 
The single-particle parameters of this cluster serve as variational 
parameters to construct a stationary point of the grand potential of
the lattice system. The stationary point is found by a crossover
procedure which allows to go continuously from an array of
disconnected clusters to the lattice system. The self-energy
is found to contain irrelevant degrees of freedom which have 
marginal impact on the grand potential and which need to be excluded
to obtain meaningful results. The obtained spectral functions are in
good agreement with experimental data.
\end{abstract} 

\maketitle
\section{Introduction}

The theoretical description of compounds containing partially
filled $3d$, $4f$ or $5f$ shells is a much-studied problem in
solid-state theory. Due to the small spatial extent of these shells
the Coulomb repulsion between the electrons in the conduction bands
formed from these shells becomes unusually strong 
and approximations which rely on a mapping of the
physical electron system onto one of fictious free particles in
a suitably constructed effective potential - as
is the case in density functional theory\cite{KohnSham} in the local density
approximation (LDA) - cannot even qualitatively
describe the resulting state. A frequently cited example are the
transition metal (TM) oxides NiO, CoO and MnO. 
Band structure calculations for the paramagnetic phase predict
these materials to be metallic while experimentally they
remain insulators well above their respective N\'eel temperature.
For NiO and MnO an insulating ground state can be obtained in the
framework of band theory by introducing antiferromagnetic order -
for CoO on the other hand even the antiferromagnetic ground state is
metallic\cite{Terakura}. From a comparison of X-ray photoemission
spectroscopy (XPS) and bremsstrahlung isochromat spectroscopy
(BIS) it was found\cite{SawatzkyAllen} that the band gap 
predicted by LDA for antiferromagnetic
is too small by a factor of $\approx 10$.
Moreover electron spectroscopy shows
that the electronic structure remains essentially unchanged
at the N\'eel temperature for both NiO\cite{Tjernberg} and
CoO\cite{Shen_CoO}. In both compounds there is practically no difference
between the electronic spectra in the antiferromagnetic and
paramagnetic phase. The same holds true for the related
compound NiS where LDA band structure calculations on the contrary 
predict that the transition to the magnetically ordered phase is
accompanied by a significant change of the electronic 
structure\cite{Fujimori_NiS}.\\
Failure to predict the insulating ground state and the
magnitude of the insulating gap is not the only
shortcoming of LDA. In valence band photoemission spectroscopy (PES)
all three oxides  NiO, CoO and MnO show a `satellite' 
at an energy of $\approx 6-8 eV$ below the valence band 
top\cite{Ohetal,Shen_CoO,Lad_Heinrich}.
The Fano-like intensity variation with photon energy 
at the TM $3p\rightarrow 3d$ threshold identifies this feature
as being due to $d^n \rightarrow d^{n-1}$ transitions\cite{Davis}. This
part of the electronic structure is not at all reproduced by band
struture calculations which on the contray
would predict the $d^n \rightarrow d^{n-1}$
transitions near the valence band top. And finally
experimental band structures
measured by angle resolved photoelectron spectroscopy (ARPES)
show that for all compounds, NiO, CoO
and MnO, the TM-derived bands near the valence band top are
almost dispersionless\cite{Shen_long,Shen_CoO,Lad_Heinrich}.
This is also in contradition to LDA calculations which
predicts band widths of around $2eV$ for the TM3d-derived bands.\\
Starting with the work of 
Hubbard\cite{Hubbard} a variety of theoretical methods have been invented to
deal with this 
problem\cite{SvaGu,Czyzyk,Bala,Manghi,Ary,Iga,Massi,nio_cpt,Kunesetal,Kunesetal_band,yinetal}.
Major progress towards a quantitative
description of $3d$ TM oxides has been made by the 
cluster method initiated
by Fujimori and 
Minami\cite{FujimoriMinami,Elp,Elp_CoO,Elp_MnO,Fujimori_MnO} 
This takes the opposite
point of view as compared to band theory, namely to abandon
translational invariance  and instead treat exactly - by means of
atomic multiplet theory\cite{Slater,Griffith} - the
Coulomb interaction in the $3d$-shell of a TM-ion in an octahedral `cage'
of nearest-neighbor oxygen atoms. The angle-integrated valence band
photoemission spectra calculated by this method are
in excellent agreement with 
experiment\cite{FujimoriMinami,Elp,Elp_CoO,Elp_MnO,Fujimori_MnO}. 
This is clear
evidence that the atomic multiplets of the partly filled $3d$-shell
- suitably modified by the crystalline electric field (CEF) -
persist in the solid and play an important part in the physics. 
On the other hand due
to its `impurity' character the cluster method can only give
$\bf{k}$-independent quantities.\\
Recently, however, ideas have been put forward to broaden the
correlated ionization and affinity states of
finite clusters into bands\cite{Oana,Senechaletal,Maier}.
A particularly elegant way to do so - the variational
cluster approximation (VCA) - has been proposed by
Potthoff\cite{PotthoffI}. Building on field-theoretical work of
Luttinger and Ward\cite{LuttingerWard} who showed that the
grand canonical potential $\Omega$ of an interacting Fermion system
is stationary with respect to variations of the electronic self-energy
$\Sigma(\omega)$, Potthoff proposed to generate trial self-energies
numerically by exact diagonalization of finite clusters and
use them in a variational scheme for $\Omega$. This amounts
to finding the best approximation to the true self-energy
of the lattice amongst the subset of
`cluster representable' ones, i.e. exact self-energies of 
finite clusters.\\
So far the VCA has been applied 
mainly to simplified systems such as
the single-band Hubbard-model\cite{PotthoffI,Dahnken} but the success of the 
cluster method for TM-oxides clearly suggests to apply the VCA also 
to a realistic
model for TM-oxides thereby using the octahedral clusters
introduced by Fujimori and Minami to generate self-energies.
Here we outline such a calculation for NiO, CoO and MnO, which all have the
rocksalt structure. For simplicity we neglect any of the lattice 
distortions observed in the actual compounds as well as 
the antiferromagnetic order at low temperature
and study the ideal rocksalt structure
in the paramagnetic phase. As already mentioned the
single-particle spectra of these compounds do not change appreciably
during the N\'eel transition so this is probably a reasonable
assumption. 
A preliminary study for NiO using the VCA has been 
published elsewhere\cite{nio_short}.\\
Using clusters containing just a single TM-ion
implies that the self-energy is site-diagonal, i.e.
${\bm k}$-independent. The
corresponding approximation thus is similar to the dynamical
mean-field (DMFT) calculations which have recently been applied to
a variety of compounds\cite{DMFTs}. The relationship between
DMFT and the VCA has been discussed in detail by
Potthoff\cite{PotthoffI} and a detailed comparison with 
recent DMFT calculations for NiO will be presented below.
\section{Hamiltonian}
\begin{figure}
\includegraphics[width=\columnwidth]{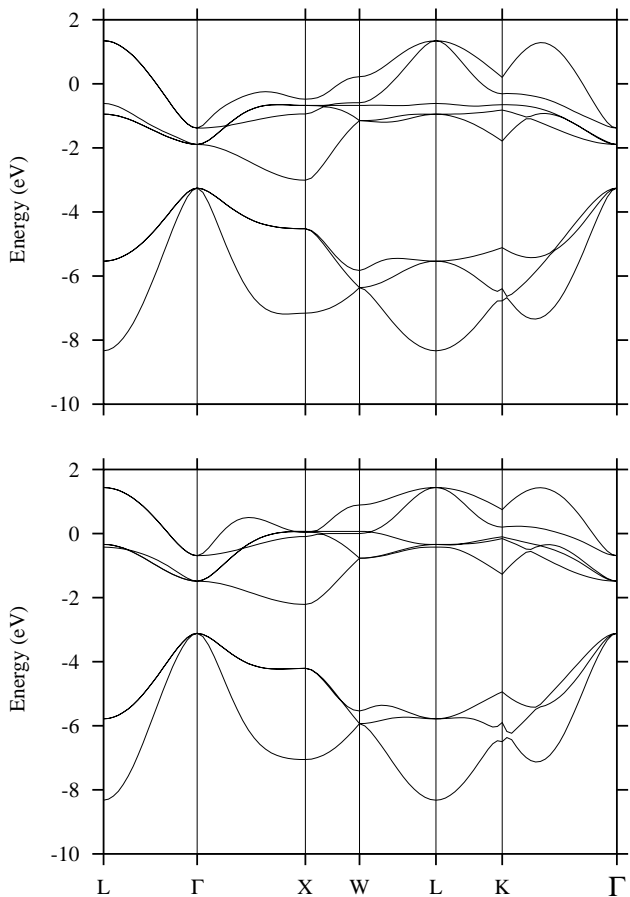}
\caption{\label{fig1}  LDA band structure (top) and LCAO-fit
(bottom) for CoO.}
\end{figure}

We start by defining the Hamiltonian which describes the correlated
TMO lattice.We use a linear combination of atomic
orbitals (LCAO) parameterization of the noninteracting 
Hamiltonian with hybridization integrals obtained from a fit to a
LDA band structure thereby closely following the procedure outlined by
Mattheiss\cite{Mattheiss}.
The parameters so obtained are listed in Table \ref{tab1}.
To give an impression about what accuracy can be expected from such a fit
Figure \ref{fig1} shows the actual LDA band structure
and the LCAO-fit for CoO.
\begin{table}[h,t]
\begin{center}
\begin{tabular}{|c|rrr|}
\hline
& NiO & CoO & MnO \\
\hline
 $(pp\sigma)$      &   0.695   &   0.627    &    0.542 \\
 $(pp\pi)$         &  -0.118   &  -0.111    &   -0.108 \\
 $(sd\sigma)$      &  -1.210   &  -1.210    &   -1.319 \\
 $(pd\sigma)$      &  -1.289   &  -1.276    &   -1.275 \\
 $(pd\pi)$         &   0.614   &   0.596    &    0.587 \\
 $(dd\sigma)$      &  -0.255   &  -0.274    &   -0.331 \\
 $(dd\pi)$         &   0.060   &   0.067    &    0.097 \\
 $\epsilon(O2s)$   & -14.000   & -14.000    &  -14.000 \\
 $\epsilon(O2p)$   &   0.000   &   0.000    &    0.000 \\
 $\epsilon(TM3d)$  &   2.822   &   3.400    &    3.899 \\
 $10Dq$            &   0.138   &   0.142    &    0.069 \\
\hline
\end{tabular}
\caption{Hybridization integrals and site-energies $\epsilon$
(in $eV$)
obtained by a LCAO fit to paramagnetic LDA band structures.
The site-energies have been shifted so as to have $\epsilon(O2p)=0$. }
\label{tab1}
\end{center}
\end{table}
Following Mattheiss\cite{Mattheiss} an $O2s$ orbital was included into the
basis set in addition to the $O2p$ and $TM3d$ orbitals.
This turned out to be crucial for a correct fit of the
dispersion of some $TM3d$-like bands along $\Gamma-X$. 
The energy of the O2s orbital has no particular impact on the
dispersion of the bands near the Fermi level
and was kept at $14eV$ below the $O2p$ energy.
The fit can be improved substantially by including also
nonvanishing overlap integrals between $O2p$ and $TM3d$
orbitals - since, however, the VCA needs well defined $TM3d$ orbitals 
to which the self-energy can be added these overlap integrals were omitted.
The hybridization element $(dd\delta)$ turned out to be meaningful
only in combination with these overlap integrals - a fit without overlap
produced a positive value of $(dd\delta)$ - and hence was set to
be zero. By and large the variation of the hybridization integrals
along the series NiO$\rightarrow$MnO is consistent with
the increasing lattice constant on one hand and the increasing
$d$-shell radius on the other. The parameters
are similar to
previous estimates in the literature\cite{FujimoriMinami,Elp}.\\
The Coulomb interaction within the $d$-shell can be written as
\begin{equation}
H_{1}=\sum_{\kappa_1,\kappa_2,\kappa_3,\kappa_4}
V_{\kappa_1,\kappa_2}^{\kappa_3,\kappa_4}\;\;
 d_{\kappa_1}^\dagger  d_{\kappa_2}^\dagger
d_{\kappa_3}^{} d_{\kappa_4}^{}.
\label{inter}
 \end{equation}
Here we have suppressed the site label $i$ and
$\kappa=(\alpha,\sigma)$ where 
$\alpha \in \{ d_{xy}, d_{xz}\dots d_{3z^2-r^2}\}$ denotes
the CEF-level. The matrix elements
$V_{\kappa_1,\kappa_2}^{\kappa_3,\kappa_4}$
can be expressed\cite{Slater,Griffith} in terms of
the $3$ Racah-parameters, $A$, $B$ and $C$.
Due to the `breathing' of the $3d$ radial wave function
these parameters should be taken to depend
on the $d$-shell occupation\cite{FujimoriMinami}. This, however, would
create an `implicit' interaction containing terms higher
than quartic in the Fermion operators and also an interaction
between $d$ and $p$-electrons. This would defeat our formalism
and we therefore do not take the dependence on $d$-shell
occupation into account.
Whereas $B$ and $C$ can be estimated from atomic Hartree-Fock wave
functions the parameter $A$ is reduced substantially from its atomic value
by solid state screening. This parameter can in principle be
obtained from density functional 
calculations\cite{Anisimov,NormanFreeman}.
In the cluster calculations for 
TM oxides\cite{FujimoriMinami,Elp,Elp_CoO,Elp_MnO,Fujimori_MnO} 
$A$ is usually treated
as an adjustable parameter and here we do the same.
Adjusting $A$ is equivalent to adjusting the `Hubbard $U$'
$=E(d^{n+1}) + E(d^{n-1}) - 2 E(d^n)$ where $E(d^n)$ denotes
the energy of a $d$-shell with $n$ electrons.
There is some ambiguity as to what exactly is to be understood
by `the energy of $d^n$' - here we follow
Refs. \cite{Elp,Elp_CoO,Elp_MnO} and take $E(d^n)$ to be the Coulomb
energy of the Hund's rule ground state of the free ion i.e.
calculated without CEF splitting. 
$E(d^n)$ - and hence $U$ - then  can be expressed in terms of the Racah
parameters\cite{Griffith}.
Another way to define a Hubbard $U$ would be to note that the
average Coulomb energy of the $d^n$ multiplets is\cite{Slater,Griffith}
\begin{equation}
E=\frac{n(n-1)}{2} (A-\frac{14}{9}B + \frac{7}{9}C)
\label{average}
\end{equation}
which would suggest to define $U_{av}= A-\frac{14}{9}B +
\frac{7}{9}C$.\\
A second parameter which of importance for
charge transfer systems\cite{ZSA} which is usually adjusted to experiment
is the $d$-level energy or equivalently the charge transfer
energy $\Delta=E(d^{n+1}\underline{L})-E(d^n)$. Expressing the 
$E(d^n)$ in terms of
Racah-parameters $\Delta$ can be expressed in terms of these and
the difference of site energies $\epsilon(TM3d) - \epsilon(O2p)$.
The values used in the present calculation are given
in Table \ref{tab2} as well. It should be noted that while
the LCAO-fit actually gives an energy for these  site energies - see Table
\ref{tab1} - these values of $\epsilon(TM3d)$ already
incorporate the Coulomb interaction between
$d$-electrons. Since in our formalism the Coulomb interaction
is described
by the Hamiltonian (\ref{inter}) one would have to subtract it off
to avoid double counting. Namely
if one considers the Hubbard $U$ as known one may estimate the
`bare' value of $\epsilon(TM3d)$ from the LDA
site-energy $\epsilon(TM3d)_{LDA}$ and the $d$-elevel
occupancies $n_d$ as\cite{Kunesetal} 
\begin{equation}
\tilde{\epsilon}_d = \epsilon(TM3d)_{LDA} - 9 U_{av} n_d.
\label{corrected}
\end{equation}
The estimates obtained in this way are also given
in Table \ref{tab2}. For NiO and CoO these corrected LDA values
are close to the adjusted parameters used in the actual
calculation. The situation is different for $MnO$ but
for this compond the $U$ obtained from
ground state energies also differs strongly from the multiplet
average $U_{av}$. The reason is the strong exchange
stabilization in the high-spin ground state of $d^5$
and one may not hope to obtain agreement between the two estimates
for the site energy either. The reason is simply that a
`Hubbard $U$' is not uniquely defined in the presence of strong
multiplet splitting.
\begin{table}[h,t]
\begin{center}
\begin{tabular}{|c|rrr|}
\hline
& NiO & CoO & MnO \\
\hline
 $A$              &   8.25  &   7.2   &    6.1  \\
 $B$              &  0.13   &   0.14  &    0.12 \\
 $C$              &  0.60   &   0.54  &    0.41 \\
 $\epsilon(TM3d)$ &  -62.0  &  -45.5  &  -23.3  \\
 $U$              &   8.38  &   7.34  &   10.65 \\
 $\Delta$         &  7.42   &  11.48  &   10.07 \\
$U_{av}$          &  8.51   &  7.40   &    6.23 \\
$n_d(LDA)$        & 0.85    &  0.76   &    0.56 \\
$\tilde{e}_d$     & -60.53  &  -45.81 &  -26.91 \\
\hline
\end{tabular}
\caption{Racah parameters, $d$-level energy, Hubbard $U$ and
charge transfer energy $\Delta$. Also given are the
`average Hubbard $U$' according to \ref{average}, the electron
number per $d$-orbital as obtained from the LDA calculation and
the estimate for the $d$-lebel energy according to \ref{corrected}.
All energies are in $eV$.}
\label{tab2}
\end{center}
\end{table}
Finally, any Coulomb interaction between electrons which are
not in the same $TM3d$-shell is neglected.
\section{Variational cluster approximation}
Having specified the stronly correlated problem under discussion
we outline the variational cluster approximation.
This is based on an expression for the  grand potential $\Omega$
of an interacting many-Fermion system derived by 
Luttinger and Ward\cite{LuttingerWard}. In a multi-band system
where the Green's function ${\bf G}({\bf k},\omega)$,
the noninteracting kinetic energy ${\bf t}({\bf k})$
and the self-energy ${\bf \Sigma}({\bf k},\omega)$
for given energy $\omega$ and momentum ${\bf k}$ are matrices of
dimension $2n\times 2n$, with $n$ the number of orbitals in the
unit cell, it reads\cite{Luttingertheorem}
\begin{eqnarray}
\Omega &=& -\frac{1}{\beta}\;\sum_{{\bf k},\nu}\; e^{\omega_\nu 0^+}
\ln\;det\;(-{\bf G}^{-1}({\bf k},\omega_\nu)+
F[{\bf \Sigma}]
\label{ydef}
\end{eqnarray}
where $\omega_\nu=(2\nu+1)\pi/\beta$ with $\beta$ the inverse temperature
are the Fermionic Matsubara
frequencies,
\begin{equation}
{\bf G}^{-1}({\bf k},\omega) =\omega + \mu - {\bf t}({\bf k})
- {\bf \Sigma}({\bf k},\omega).
\label{gdef}
\end{equation}
with $\mu$ the chemical potential
and the functional $F[{\bf \Sigma}]$ is the Legendre transform of the
Luttinger-Ward functional $\Phi[{\bf G}]$.
The latter is defined\cite{LuttingerWard} 
as the sum of all closed linked skeleton diagrams with the non-interacting
Green's functions replaced by the full Green's functions.
A nonperturbative derivation of a functional with the same properties
as $\Phi$ has recently been given by Potthoff\cite{Nonperturbative}.
$\Phi[{\bf G}]$ is the generating functional of the self-energy
${\bf \Sigma}$ i.e.
\begin{equation}
 \frac{1}{\beta}\; { \Sigma}_{ij}({\bf k},\omega_\nu)=
\frac{\partial \Phi}{\partial  G_{ji}({\bf k},\omega_\nu)},
\end{equation}
and $F[{\bf \Sigma}]$ is obtained by Legendre-transform to
eliminate ${\bf G}$ in favour of ${\bf \Sigma}$:
\[
F[{\bf \Sigma}]=\Phi[{\bf G}] -\frac{1}{\beta}
\sum_{{\bf k},\nu} trace\left(({\bf G}({\bf k},\omega_\nu){\bf \Sigma}({\bf
  k},\omega_\nu)\right).
\]
By virtue of being a Legendre transform it obeys
\begin{equation}
\frac{1}{\beta}\; { G}_{ij}({\bf k},\omega_\nu)=
-\frac{\partial F}{\partial \Sigma_{ji}({\bf k},\omega_\nu)},
\end{equation}
and using the identity
\begin{equation}
\frac{\partial }{\partial A_{ij}} \ln\;det\;A = (A^{-1})_{ji}
\label{inverse}
\end{equation}
which holds for any matrix $A$ with $det\;A\ne 0$
as well as the Dyson equation (\ref{gdef}) we find
that $\Omega$ is stationary with respect to variations of
${\bf\Sigma}$:
\begin{equation}
\frac{\partial \Omega}{\partial \Sigma_{ij}({\bf k},\omega_\nu)} = 0.
\label{stationary}
\end{equation}
The crucial obstacle in exploiting this stationarity property 
in a variational scheme for the self-energy
${\bf \Sigma}$ is the evaluation of the functional $F[{\bf \Sigma}]$ for
a given `trial ${\bf \Sigma}$'. Potthoff's
solution\cite{PotthoffI} makes use of the fact that
just like $\Phi[{\bf G}]$, $F[{\bf \Sigma}]$ 
has no explicit dependence on the single-particle terms
of $H$ and therefore
is the same functional of ${\bf \Sigma}$ for any two systems
with the {\em same interaction part} of the Hamiltonian.
This is easily seen from the diagrammatic representation
of $\Phi[{\bf G}]$ because the expression associated with
a given Feynman diagram involves only the interaction matrix elements
and the Green's function itself. \\
In the VCA this independence of $F[{\bf \Sigma}]$  on the
single-particle terms of the Hamiltonian is used to construct
trial self-energies by exact diagonalization of
finite clusters and thereby obtain the exact numerical value
of $F[{\bf \Sigma}]$. In a first step one chooses a so-called
reference system, which has the same interaction part
as the lattice problem under study but consists
of disconnected finite clusters. If the interaction terms are
short ranged - which is the reason for keeping
only the Coulomb interaction between electrons in the same $d$-shell -
this can 
can always be achieved by suitable choice of the single-particle terms.
The disconneted finite clusters of the reference system
then are solved by exact
diagonalization, which gives the eigenenergies $\epsilon_\nu$
and wave functions $|\Phi_\nu\rangle$ 
for all particle numbers in the cluster. Of course this sets
some limit on the size of the clusters. Next, the
Green's function $\tilde{\bf G}(\omega)$ and grand
potential $\tilde{\Omega}$ of the reference system
are calculated numerically and equation (\ref{ydef}) 
is reverted to express the exact 
numerical value of $F[{\bf \Sigma}]$ in terms of these.
This simply means that the summation of infinitely many Feynman diagrams
and Legendre transform is done implicitely in the course of the
exact diagonalization of the reference system.
Then, the self-energy ${\bf \Sigma}(\omega)$ of the reference system -
which is readily extracted from the Dyson equation for 
$\tilde{\bf G}(\omega)$ -
can be used as a trial self-energy
for the lattice system. Thereby the numerical value
of $F[{\bf \Sigma}]$ calculated in the cluster
is simply inserted into the Luttinger-Ward formula(\ref{ydef}) for the
grand potential of the physical (i.e. lattice) system.
The variation of ${\bf \Sigma}(\omega)$ is performed by varying the
single-electron parameters - such as hybridization integrals or
site-energies - of the reference system. \\
In applying this procedure one frequently has to evaluate expressions
of the type (the momentum $\bm{k}$ is suppressed for brevity)
\begin{equation}
S = -\frac{1}{\beta} \sum_\nu e^{\omega_\nu 0^+}\;
\ln\;det\;(-\bm{G}^{-1}(\omega_\nu)).
\label{sdef}
\end{equation}
To evaluate this we closely follow Luttinger and Ward\cite{LuttingerWard}
and first convert the sum into a contour integral
\begin{equation}
-\frac{1}{\beta} \sum_\nu g(\omega_\nu) \rightarrow \frac{1}{2\pi i}
\int_{C_0} f(\omega) g(\omega) d\omega
\label{contour}
\end{equation}
where $f(\omega)$ is the Fermi function and $C_0$ is the standard
contour encircling the singularities of the Fermi function in
counter-clock-wise direction. 
Next we deform the contour so that it encircles the singularities of
the logarithm, which are all located on the real axis (see the Appendix).
Following Luttinger and Ward we then use
\begin{equation}
f(\omega)= -\frac{1}{\beta} \frac{d}{d\omega} \log(1 + e^{-\beta
  \omega})
\label{luttytrick}
\end{equation}
and integrate by parts. Using (\ref{inverse}) we thus obtain
\begin{eqnarray}
S&=& \frac{1}{2\pi\beta i} \int_{C}
d\omega\; \log(1 + e^{-\beta \omega})\nonumber \\
&&\;\;\;\;\;\;\;\;\;\;\;\;\;\;\;\;\;\;
trace\;[ (1 -\frac{d \bm{\Sigma}(\omega)}{d\omega})\bm{G}(\omega)]
\end{eqnarray}
where $C$ is a contour that encircles the singularities
of $trace$ in clockwise fashion. This can now be evaluated
by numerical contour integration.
To derive the expression given by Potthoff we note the
alternative expression
\[
S=\frac{-1}{2\pi\beta i} \int_{C}d\omega \log(1 + e^{-\beta \omega})\;
\sum_i \frac{1}{\lambda_i(\omega)}\frac{\partial
  \lambda_i(\omega)}{\partial \omega}
\]
where $\lambda_i(\omega)$ are the eigenvalues of $\bm{G}(\omega)$.
There are two types of singularities of this expression:\\
a) zeroes of an eigenvalue, i.e.
\[
\lambda_(\omega)\approx a_{\nu}
(\omega - \zeta_{\nu}) \;\rightarrow\;
\frac{1}{\lambda(\omega)}\frac{\partial
  \lambda(\omega)}{\partial \omega} = \frac{1}{\omega - \zeta_{\nu}}
\]
b) singularities of an eigenvalue, i.e.
\[\lambda(\omega)\approx \frac{b_{\mu}}{\omega - \eta_{\mu}}
\;\rightarrow\;
\frac{1}{\lambda(\omega)}\frac{\partial
  \lambda(\omega)}{\partial \omega} =-\frac{1}{\omega - \eta_{\mu}}
\]
Whence
\[
S = -\frac{1}{\beta}\left( \sum_\mu \; 
\log(1 + e^{-\beta \eta_\mu}) - 
\sum_\nu \log(1 + e^{-\beta \zeta_\nu}) \right),
\]
i.e. the expression derived by Potthoff\cite{PotthoffI}.\\
\begin{figure}
\includegraphics[width=\columnwidth]{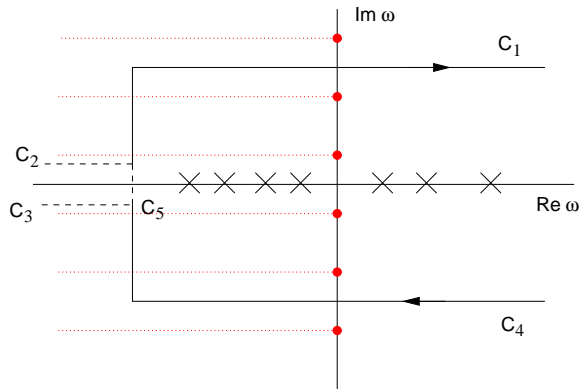}
\caption{\label{fig2} Integration contour for the numerical
evaluation of $S$. The dots on the imaginary axis denote
the Matsubara frequencies, the dotted lines are
branch cuts of $\log(1 + e^{-\beta  \omega})$. Crosses denote
poles of the Greens function.}
\end{figure}
For the numerical evaluation of $S$ a slightly different
procedure is more convenient. For Matsubara
frequencies $\omega_\nu$ with $|\nu|\le\nu_{max}$ the respective terms
in the sum (\ref{sdef}) are evaluated
directly by computing the eigenvalues
of $\bm{G}^{-1}(\omega)$. For $|\nu|>\nu_{max}$ we switch to a contour
integral using (\ref{contour}) and deform the
integration contour as indicated in Figure \ref{fig2}.
Along $C_1$ and $C_4$ the integral is evaluated numerically
again by calculating the eigenvalues of $\bm{G}^{-1}(\omega)$.
Along the positive real axis the integrand thereby
is cut off by the 
Fermi function.
Along $C_2$ and $C_3$ we integrate by parts using again
(\ref{luttytrick}) and deform the two pieces into
the short piece $C_5$ - which is possible
because the contour encloses no more singularities
of the integrand.
It is not possible to integrate by parts along $C_1$ and $C_4$
because these contours cross the lines - indicated by dashed lines
in Figure \ref{fig2} - where 
$\log(1 + e^{-\beta  \omega})$ has branch cuts.
The advantage of this procedure is that the resulting
integration contour $C_1 - C_5 -C_4$ can be kept far from
the singularities of the Greens function and self-energy
on the real axis so that the integrand will always be a smooth function
and a numerical integration with relatively few
mesh points (of order $10^3$) gives accurate results.
\begin{figure}
\includegraphics[width=\columnwidth]{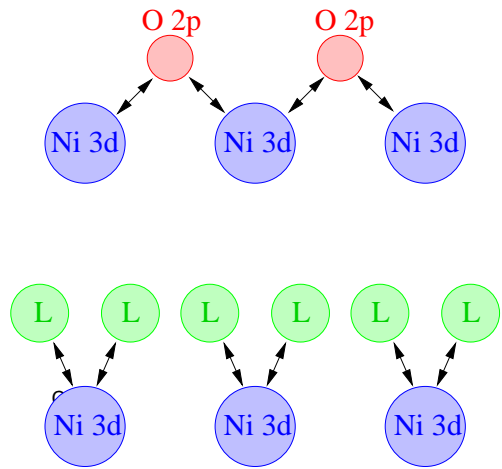}
\caption{\label{fig3} Physical system (top) and reference
system (bottom). Arrows indicate nonvanishing
hybridization. While the physical system is a true
lattice, the reference system is an array of disconnected
clusters. The Coulomb interaction between electrons
in the TM $d$-shell is the same for both systems, however. }
\end{figure}
\section{Reference System}
\begin{figure}
\includegraphics[width=\columnwidth]{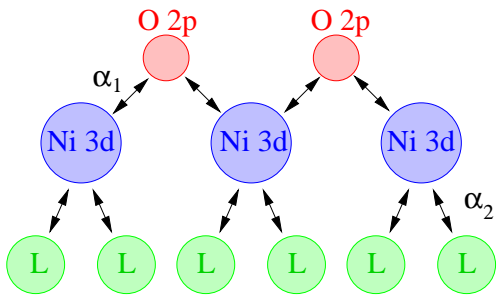}
\caption{\label{fig4} `Hybrid system' used to find the stationary
  point of $\Omega$.}
\end{figure}

\begin{figure}
\includegraphics[width=\columnwidth]{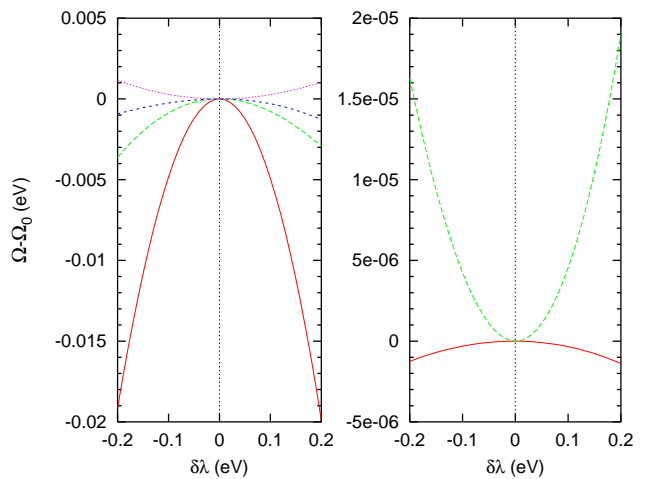}
\caption{\label{fig5} Scans through the stationary point of $CoO$. 
The normalized
eigenvectors ${\bf v}_i$
of $\partial^2 \Omega /\partial \lambda_i \partial
\lambda_j$ were muliplied by $\lambda_i$.}
\end{figure}

Given the excellent results
obtained by the cluster method for angle-integrated 
spectra\cite{FujimoriMinami,Elp,Elp_CoO,Elp_MnO,Fujimori_MnO} it
seems natural to use clusters which are equivalent to
TMO$_6$ octahedra as reference system.
More precisely we choose a reference system where each $TM3d$ 
orbital $d_\alpha$ 
is coupled to one `ligand' orbital $L_\alpha$ with these ligands in turn
decoupled from each other and the interaction
within the $d$-shell given by (\ref{inter}). The reference system thus is
equivalent to an array of non-overlapping identical
TMO$_6$ clusters where each ligand $L_\alpha$
corresponds to the unique linear combination of $O2p$
orbitals on the six nearest $O$ neighbors of a given TM atom which hybridizes
with the $TM3d_\alpha$ orbital - see Figure \ref{fig3}. 
We write the single-particle Hamiltonian for a TML$_5$ cluster as
\begin{eqnarray}
H_{single} &=& \sum_{\alpha,\sigma}\;V(\alpha)\left(\;d_{\alpha,\sigma}^{\dagger}
L_{\alpha,\sigma}^{} + H.c.\;\right) \nonumber \\
&& + \sum_{\alpha,\sigma}
\left(E({\alpha})\;d_{\alpha,\sigma}^{\dagger}
d_{\alpha,\sigma}^{} + e({\alpha}) \;L_{\alpha,\sigma}^{\dagger}
L_{\alpha,\sigma}^{}\right).\nonumber \\
\label{heff}
\end{eqnarray}
The variational parameters thus are the hopping matrix elements
$(V_\alpha)$, the ligand energies $e(\alpha)$ and
the $d$-level energies $E(\alpha)$ with
$\alpha \in \{e_g, t_{2g}\}$ - in total we thus
have $6$ parameters.
It is convenient to rewrite the site-energies in the reference system
as follows
\begin{eqnarray}
E(e_g)    &=& \epsilon_0 - \epsilon_1 + 3\epsilon_2/5 + \epsilon_d,\\
E(t_{2g}) &=& \epsilon_0 - \epsilon_1 - 2\epsilon_2/5 + \epsilon_d,\\
e(e_g)    &=& \epsilon_0 + \epsilon_1 + 3\epsilon_3/5,\\
e(t_{2g}) &=& \epsilon_0 + \epsilon_1 - 2\epsilon_3/5.
\label{rewrite}
\end{eqnarray}
As shown by Aichhorn {\em et al.}\cite{aichhornetal} optimization of the
`center of gravity' $\epsilon_0$ ensures that the electron number
obtained by differentiating $\Omega$ with respect to $\mu$
is equal to the result obtained by integrating the spectral function.\\
The search for the stationary point of a function of
$6$ variables $\lambda_i$ is a difficult task - even more so because
the stationary point is not a global minimum or maximum and is in fact
not even a local extremum but a saddle point (see below).
This problem has motivated the search for functionals other
than $\Omega[\Sigma]$ which take an extremum value
at the physical self-energy\cite{Nevidomskyy}.
We can solve this problem in the following way, however:
if we have a set of parameters which is sufficiently close to the 
stationary point we can evaluate the derivatives 
$\partial \Omega /\partial \lambda_i$ and
$\partial^2 \Omega /\partial \lambda_i \partial \lambda_j$ numerically
and use the Newton method to find the point
where $\partial \Omega /\partial \lambda_i=0$.
Next, instead of the true lattice system we choose our
`physical system' to be 
a `hybrid system' which contains both, the $O2p$-lattice and the
Ligands for each TM-ion  see Figure \ref{fig4}.
We take the $TM-O$ and $TM-TM$ hybridization
to be multiplied by $\alpha_1$, the $TM-L$ hybridization by
$\alpha_2$. For $\alpha_1=0$ and $\alpha_2=1$ we therefore have the
reference system itself plus a decoupled $O2p$-lattice.
For this system the exact stationary point is known - namely the 
parameters of the reference system itself. On the other hand, for
$\alpha_1=1$ and $\alpha_2=0$ we have the physical lattice system
plus the decoupled and hence irrelevant ligands and a solution to this
system is a solution to the lattice system itself. In this way
we can go continuously
from an exactly solvable system to the physical lattice system.
In practice
we vary the parameters $\alpha_1,\alpha_2$ in steps of $0.1$
and start the Newton method using the stationary values of the preceding
step as initial values. When combined with a simple extrapolation
procedure this yields the stationary point of the lattice
in $\approx 10$ steps with $\le 2$ Newton-iterations in each step.
Obviously such a crossover procedure can be formulated for
other applications of the VCA as well.\\
We next discuss a second important technical point.
Having found the stationary point we can calculate the
matrix of second derivatives 
$\partial^2 \Omega /\partial \lambda_i \partial \lambda_j$ 
and diagonalize it. 
Figure \ref{fig5} shows a scan of $\Omega$ through the
stationary point of $CoO$ along the principal axes so obtained.
There are two important things to recognize. First, the stationary
point is a saddle point but the above crossover procedure
had no problems to find it. Second there are certain
directions in parameter space where $\Omega$ shows only an extremely
weak variation. This turned out to be true in all other
cases studied as well. This weak variation may either stem from a 
near-invariance of the self-energy under changes of the cluster-parameters
or the change of the self-energy is appreciable but irrelevant
in that it does not change the lattice $\Omega$. 
The presence of such `nearly invariant lines' in parameter space
clearly is undesirable in that it may induce numerical instabilities. 
It may happen that
small changes of $\Omega$ due to e.g. a slightly wrong LCAO band structure
or even numerical inaccuracies may
drive the stationary point along these lines in parameter space
to compensate for them. 
To simplify matters the number of parameters therefore
was reduced. Inspection of the eigenvectors associated
with the `nearly invariant lines'
showed, that these were predominantly combinations of the
parameters $\epsilon_1$ and $\epsilon_3$ in (\ref{rewrite}).
These parameters have
almost no influence on $\Omega$ and hence 
were not subject to variation. The parameter
$\epsilon_1$ was set equal to zero.
In the cluster calculation the value for $\epsilon_3$
would be $2(pp\sigma)-2(pp\pi)$\cite{Elp} - for simplicity 
the value $\epsilon_3=1.4eV$ was used for all three compounds.\\
Including $\epsilon_1$ and $\epsilon_3$
into the set of parameters to be optimized actually turned out to give
unsatisfactory results for the
single particle spectrum. $\epsilon_1$ tended to take on large
positive values whereas
$\epsilon_3$ usually took large {\em negative} values.
As a net effect this produced
spurious photoemission peaks with very small spectral weight
which were split off by one or two $eV$ from the remainder of the
photoemission spectrum resulting in too small gaps and poor agreement
with experiment. Clearly this is a feature of the variational cluster 
approximation which needs to be clarified.
It should be noted that reducing the number of parameters which are 
optimized simply
amounts to restricting the space of trial self-energies.
Since optimization of these parameters hardly changes $\Omega$
this is similar to restricting the degrees of freedom
in a trial wave-function to the most relevant ones.
It then seems that an `overoptimization' of parameters leads to poor
results but on the other hand the inclusion of irrelevant degrees
of freedom into a variational wave function may also
be detrimental for properties of the wave function other than
the ground state energy.\\
Finally, NiO turned out to be a special case.
Since the ground state of $d^8$
in cubic symmetry is $t_{2g}^6e_g^2$ the hopping integral
$V(t_{2g})$ has practically no impact on $\Omega$ 
because it connects filled orbitals. In fact, derivatives of $\Omega$
with respect to $V(t_{2g})$ turned out to be of order
$10^{-10}$, i.e. well beyond the numerical accuracy of the
whole procedure. $V(t_{2g})$ was therefore kept at
$2(pd\pi)$ which again is the value expected in the cluster
calculation.
In a previous VCA-study of NiO\cite{nio_short} a different approach was chosen.
There the hopping integral $V(t_{2g})$ was set set equal to zero.
This implies that the $t_{2g}$-like ligands are irrelevant alltogether
and can be discarded from the reference system, so that also
the paramaters $E(t_{2g})$ and $e(t_{2g})$ play no more role.
Although slightly different LCAO- and Racah-parameters
were used in this calculation the results obtained in 
Ref. \cite{nio_short}
are very similar to the ones in the present study, in particular the
bands in the valence band top are essentially identical.

\section{Results}
\begin{figure}
\includegraphics[width=\columnwidth]{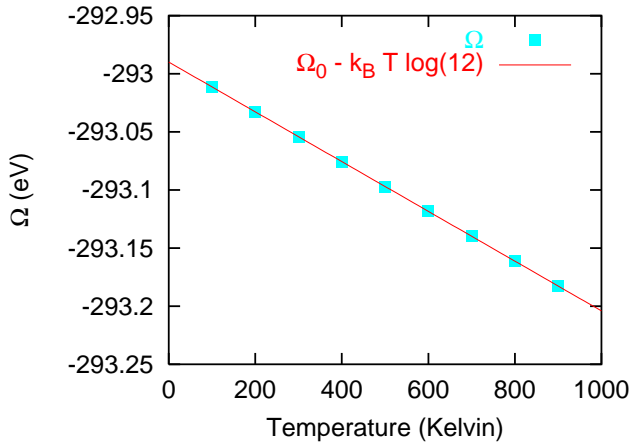}
\caption{\label{fig6} Grand potential $\Omega(T)$
obtained by the VCA for CoO.}
\end{figure}

The parameters at the stationary point of $\Omega$ 
have practically no dependence on temperature.
Figure \ref{fig6} shows the temperature dependence of
$\Omega$ for CoO.
\begin{figure}
\includegraphics[width=\columnwidth]{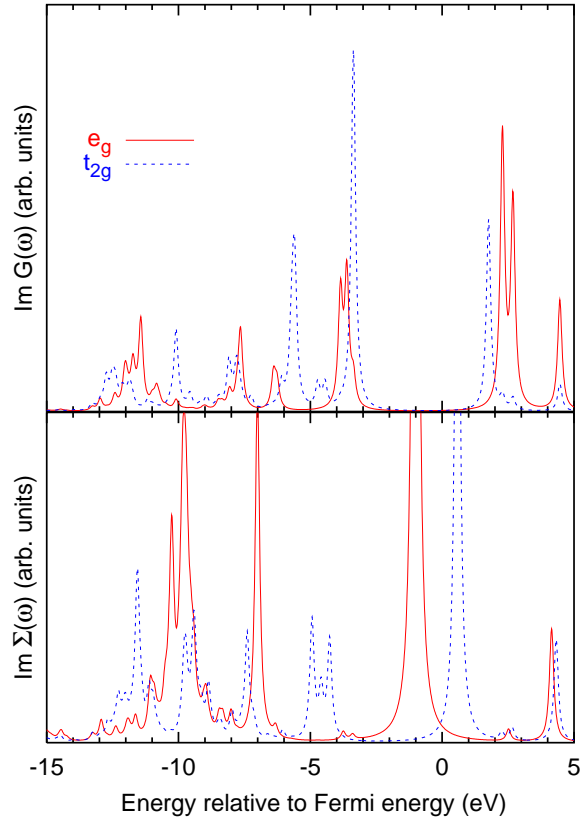}
\caption{\label{fig7} Imaginary part of the
Green's function (top) and self-energy
(bottom) for the reference CoO$_6$ cluster. The imaginary part
of the frequency is $-0.1eV$. 
Parameter values correspond to the stationary point at 150$K$.}
\end{figure}
This
can be fitted very well by $\Omega(T) = \Omega_0 - k_B T \log(12)$.
The second term thereby is the entropy due to the
degeneracy of the $^4T_{1g}$ ground state
of $d^7$ in cubic symmetry. In a system with a wide gap
this is the expected behaviour of $\Omega$. This is clearly a trivial
result but it should be noted that for the discussion of a 
phase transition to a
magnetically or orbitally ordered state the correct description
of this entropy is important because this competes
with the energy gain due to ordering.\\
Next we consider the resulting self-energy.
Figure \ref{fig7} shows the spectral density of the CoO$_6$ cluster
and the self-energy at the stationary point
for $T=150$ Kelvin. Due to the cubic symmetry of the cluster
only the diagonal elements of the self-energy are non-vanishing and these
are identical between all $e_g$ and $t_{2g}$ orbitals, respectively.
Luttinger has shown\cite{Luttingerself}
that the self energy has a spectral
representation of the form 
\begin{equation}
{\bf \Sigma}(\omega) = {\bf \eta} +
\sum_\nu \frac{{\bf S}_\nu}{\omega -\zeta_\nu}
\label{selfrep}
\end{equation}
where the real matrix ${\eta}$ is actually the
Hartree-Fock potential and the poles $\zeta_\nu$ are all on the real
axis.
The spectral density of the cluster has a well-defined
gap around $\omega=0$ between a charge-transfer peak
and the upper Hubbard band.
The self-energy for both $e_g$ and $t_{2g}$ electrons
has a strong central peak (indicating a pole $\zeta_\nu$
with large residuum $S_\nu$) in this gap.
Using the Dyson-equation it is easy to convince
oneself that such a strong peak in the self-energy indeed
`pushes open' a gap in the pole structure of the Green's function.
Several other prominent peaks create additional
gaps in the spectral density and thus split off the
satellite below $-9eV$.
In addition there are many small peaks near the top of the valence
band. Since the poles of the Green's function are
`sandwiched' between the poles of the self-energy we thus expect
a large number of 3d-derived bands with very small dispersion in this
energy range.\\
\begin{figure}
\includegraphics[width=\columnwidth]{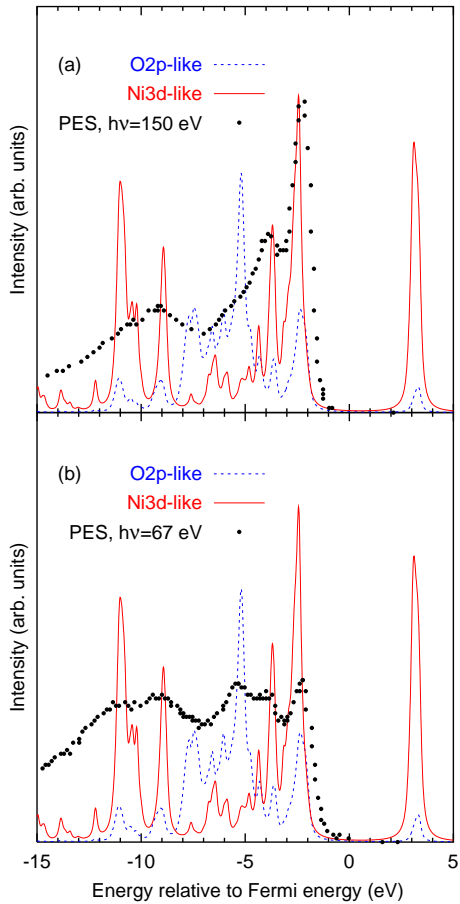}
\caption{\label{fig8} Single particle spectral densities
obtained by VCA for NiO
(${\bf k}$-integrated with 110 ${\bf k}$-points in the irreducible wedge
 of the Brillouin zone)
compared to angle-integrated valence band photoemission data\cite{Ohetal}.}
\end{figure}
\begin{figure}
\includegraphics[width=\columnwidth]{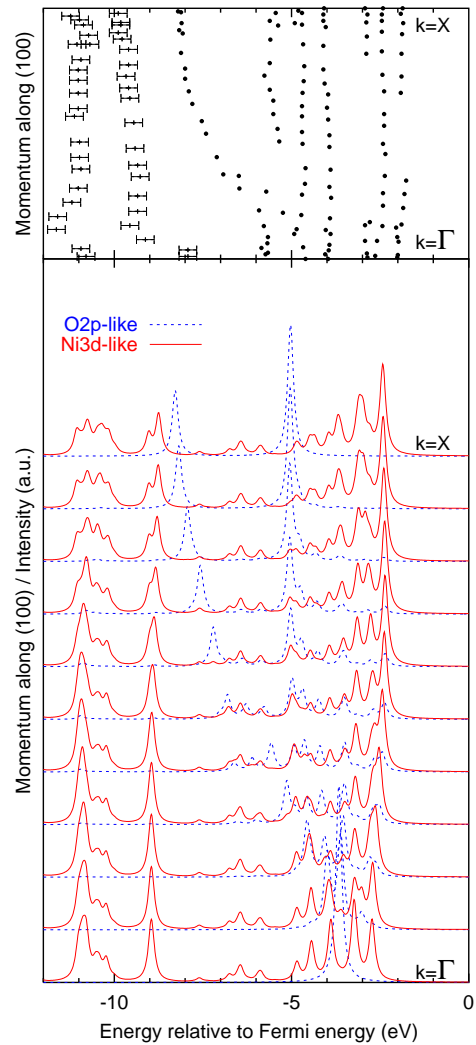}
\caption{\label{fig9} Top: Dispersion of experimental bands
measured by ARPES\cite{Shen_long} in NiO. Bands in the satellite region are
given with error bars due to their strong broadening.
Bottom: $\bm{k}$-dependent
spectral function for momenta along $\Gamma-X$ in NiO. Lorentzian broadening
$0.05\;eV$, $d$-like weight is multiplied by factor of $2$.}
\end{figure}
\begin{figure}
\includegraphics[width=\columnwidth]{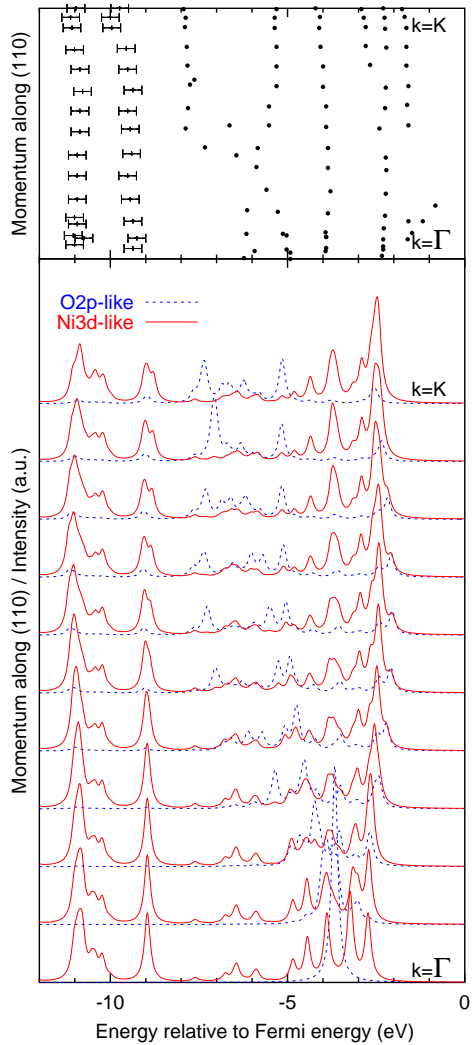}
\caption{\label{fig10} Same as Figure \ref{fig9} but for momenta
along the $(110)$ direction ($\Gamma-K$).}
\end{figure}
Next, we proceed to a comparison of  $\bf{G}(\bf{k},\omega)$ to
experiment and begin with NiO.
Figure \ref{fig8} compares $\bm{k}$-integrated
spectral densities at $T=150$ Kelvin to
angle integrated valence band photoemission spectra
taken by Oh {\em et al.}\cite{Ohetal} 
at two different photon energies.
It is known\cite{Eastman_Freeouf} that with decreasing photon energy the
intensity of $O2p$ derived states increases relative to that of
$TM3d$ derived states - the change of the spectra with photon
energy thus allows to draw conclusions about the character
of individual peaks.
Moreover, final states with $d^{n-1}$ character are 
enhanced at a photon energy just above the $TM3p\rightarrow
TM3d$ absorption threshold so that such energies are particularly
suited to identify this type of final state.
Accordingly, at $h\nu=150eV$ the experimental spectrum 
mostly resembles the $d$-like spectral density, whereas 
at $h\nu=67eV $ the states at the valence band top
are anti-resonantly supressed - whence $O2p$-derived features become 
more clearly visible - and the 
`satellite' at $-10eV$ is resonantly enhanced. 
Figures \ref{fig9}  and \ref{fig10} compare the $\bm{k}$-resolved 
spectral function
for momenta along $(100)$ $(\Gamma \rightarrow X)$ 
and $(110)$ $(\Gamma \rightarrow K)$ 
to the experimental band dispersion by Shen {\em et al.}\cite{Shen_long}.\\
The spectral density has gap of approximately $4eV$ around
the chemical potential. This is consistent
with experiment\cite{SawatzkyAllen} but has of course been achieved by
the choice of $A$ and $\Delta$.
At the top of the photoemission spectrum, $E<0$,
there is a high-intensity band complex
at binding energies between $\approx -3.5\;eV $
and $\approx -2\;eV $, which was shown to
consist of several sub-peaks by Shen {\em et al.}\cite{Shen_long}.
These authors did not actually resolve the dispersion of the
individual sub-peaks
although the data seem to indicate a weak overall `upward'
dispersion as one moves $\Gamma \rightarrow X$
which would be consistent with theory. 
Proceeding to more negative binding energy  the experimental
band structure shows a gap of
$\approx 1eV$ and then a group of dispersionless bands
between $-6eV$ and $-4eV$. This is bounded from below by a weakly
dispersive band which resembles one of the $O2p$ derived
bands. In the angle-integrated spectrum, Figure \ref{fig8}a,
the topmost of these dispersionless bands produces the `shoulder' at 
$-4eV $. The gap between the topmost band complex
and the group of dispersionless bands in the theoretical spectra
is not as wide as in experiment
but there are clearly 
several dispersionless bands in approximately the right energy range.
The agreement would be very good if the peaks at $\approx -3.5eV$
in Figures \ref{fig9}  and \ref{fig10} were shifted by $\approx 0.5eV$
to more negative binding energy.
In addition, the $O2p$ derived band can be seen clearly.
As can be seen from the band structure in Figure \ref{fig1}
this band actually has a saddle point at $X$ - this gives rise to
a van Hove-singularity in the angle-integrated spectrum which matches
very well the peak at $\approx -5.5eV$ in Figure \ref{fig8}b.
The sole strongly dispersive feature
in the spectrum, namely an $O2p$-derived band at binding
energies between $-6eV\rightarrow -9eV$ is again well reproduced by theory.
Finally the `satellite' at binding
energies $-8eV\rightarrow -12eV$ consists of at least two sub-peaks
as can be seen in Figure \ref{fig8}b and also in the ARPES data. \\
\begin{figure}
\includegraphics[width=\columnwidth]{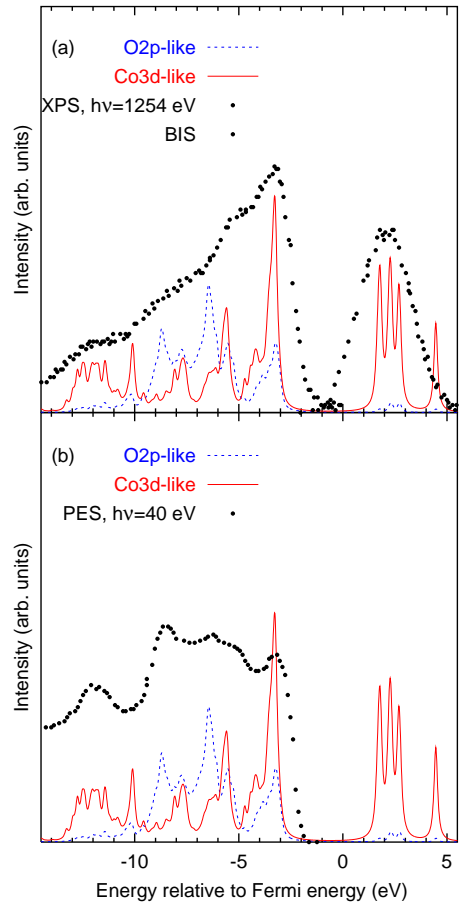}
\caption{\label{fig11} ${\bf k}$-integrated
Single particle spectral densities
obtained by VCA for CoO compared to valence band photoemission data.
Experimental data in (a) are from Ref. \cite{Elp_CoO},
in (b)  from Ref. \cite{Shen_CoO}.}
\end{figure}
\begin{figure}
\includegraphics[width=\columnwidth]{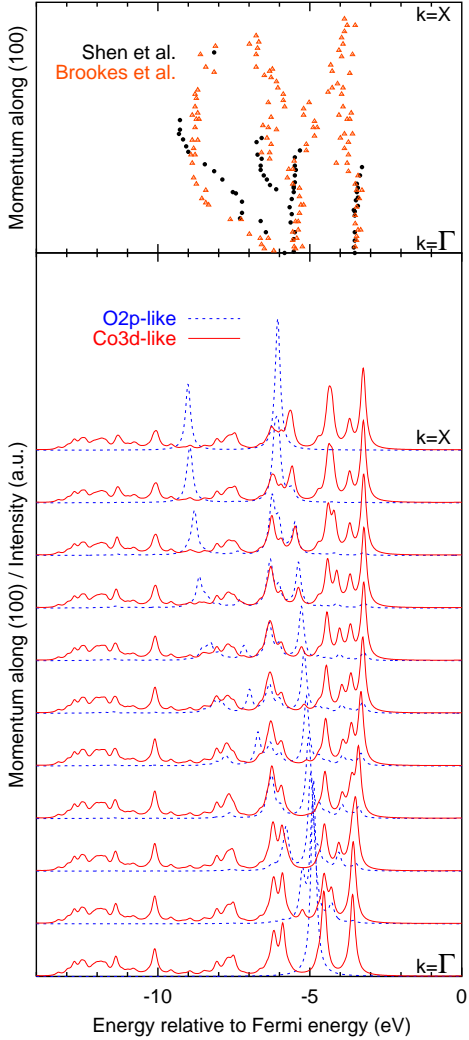}
\caption{\label{fig12} Top: Experimental band structure along $(100)$
as seen in ARPES by Shen {\em et al}\cite{Shen_CoO}
and by Brookes {\em et al}\cite{Brookesetal} in CoO. 
Bottom: $\bm{k}$-dependent spectral function for momenta along 
$\Gamma-X$ for CoO. Lorentzian broadening
$0.05\;eV$, $d$-like weight is multiplied by factor of $2$.}
\end{figure}
Next we consider CoO. Figure \ref{fig11} compares the angle integrated
spectra at different photon energies and the $ {\bf k}$-integrated
spectral function obatined by the VCA, Figure \ref{fig12}
shows the dispersion along $\Gamma\rightarrow X$ and
ARPES data from Shen {\em et al.}\cite{Shen_CoO} and
Brookes {\em et al.}\cite{Brookesetal}.
The XPS spectrum for CoO starts out with a prominent peak at
$-3eV$ followed by three `humps' at $-5eV$, $-8eV$ and $-12eV$.
The VCA gives peaks of $d$-like spectral weight
at roughly these energies although the peak at $-5eV$
is at slightly too negative energy. The PES spectrum at
$40eV$ shows additional peaks at $-6.5eV$ and $-9eV$ which were
interpreted as O2p-derived by Shen {\em et al.}\cite{Shen_CoO}.
These peaks are also reproduced by the VCA.
A little more problematic is the angle resolved spectrum. Along
(100) the VCA predicts a split peak at the the top of the
valence band at $-4eV$. This splitting is not seen in 
experiment - on the other hand, the spectra were taken at low photon
energy where the Co3d states have relatively small weight.
There is another $d$-derived band at $-6eV$
which corresponds to the second `hump' in the
angle integrated spectrum. This is crossed by and mixes
with one of the O2p derived bands, which start at $\Gamma$ at
an energy of $-5eV$. The presence of these two crossing
bands may explain the `wiggly' nature of the bands
observed experimentally in this energy range. The presence of more
than one band and a possible crossing between these is clearly seen
in the data of Shen {\em et al.}. Finally there is the
strongly dispersive O2p derived band at energies of around
$-8eV$. Surprisingly the experimental dispersions for this
band differ somewhat - this may be due to the crossing of this
band with the dispersionless Co3d derived band at $-7.5eV$
which leads to a hybridization gap in the O2p derived band.\\
\begin{figure}
\includegraphics[width=\columnwidth]{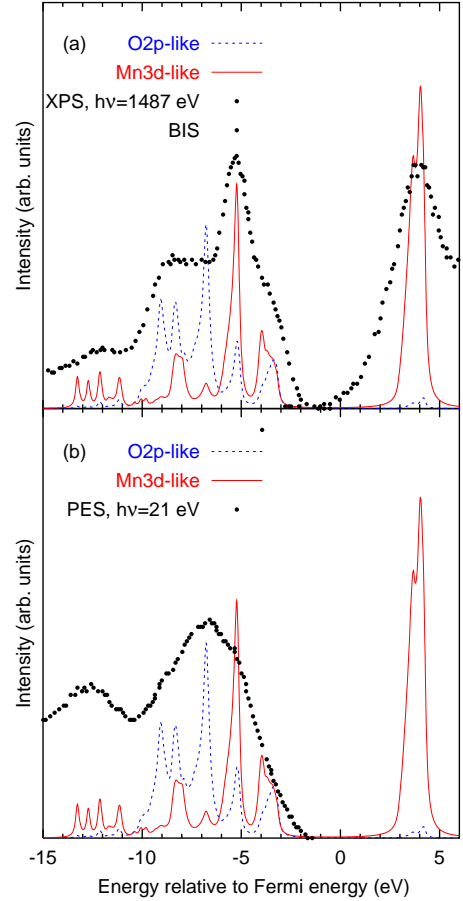}
\caption{\label{fig13} Single particle spectral densities
obtained by VCA for MnO
compared to valence band photoemission data. Experimental data
in (a) from Ref. \cite{Elp_MnO}, in (b) from Ref. \cite{Fujimori_MnO}.}
\end{figure}
\begin{figure}
\includegraphics[width=\columnwidth]{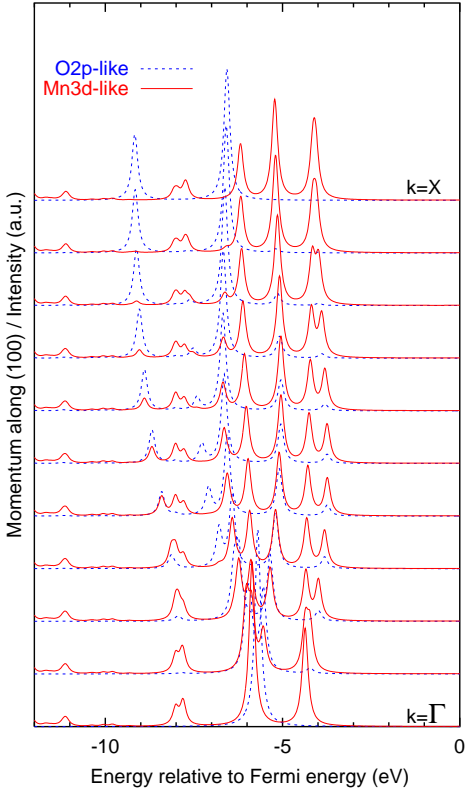}
\caption{\label{fig14} 
Spectral function for momenta along $\Gamma-X$ in MnO. 
Lorentzian broadening
$0.05\;eV$, $d$-like weight is multiplied by factor of $2$.
There are no ARPES results available.}
\end{figure}
Finally we consider MnO. Figure \ref{fig13} shows the angle integrated
photoemisison spectra compared to the result from the VCA.
At high photon energy the experimental spectrum matches
well the $d$-derived density of states. The single-particle gap
and the structure of the valence band spectrum are reproduced well.
At a photon energy of $20eV$ the intense peak at $-5.5eV$
almost disappears and another large peak at $-6.5eV$
appears, which accordingly must have O2p character.
In the theoretical spectra this is reproduced well,
the peak at $-6.5eV$ again is due to a van-Hove singularity
at $X$. Figure \ref{fig14} shows the
${\bf k}$-resolved spectrum along $(100)$. 
Lad and Heinrich\cite{Lad_Heinrich} performed ARPES measurements
on MnO but did not perform any band mapping due to the
broad nature of peaks so that an experimental dispersion unfortunately
is not available.\\
\begin{figure}
\includegraphics[width=\columnwidth]{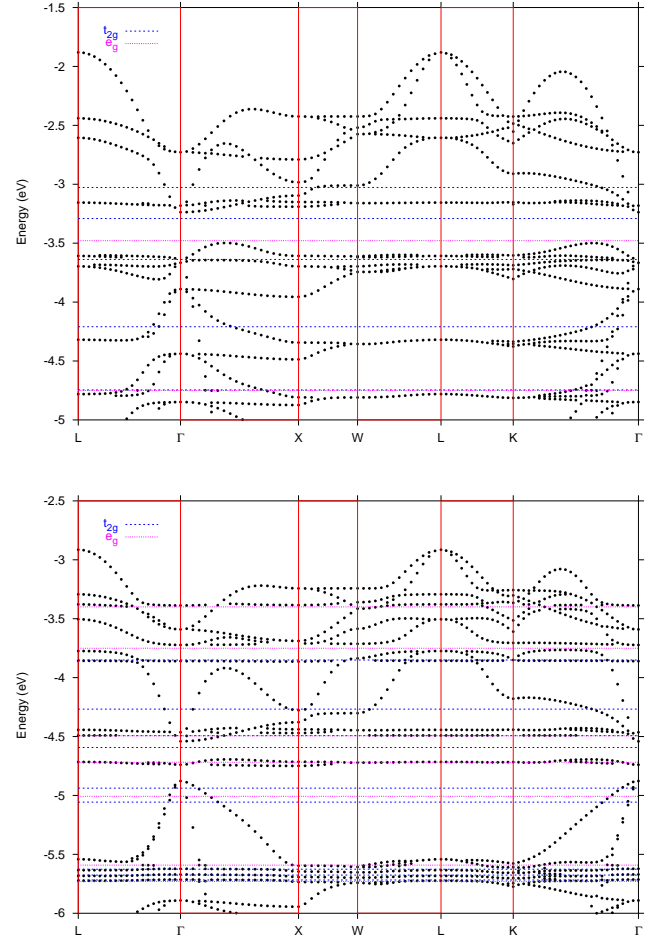}
\caption{\label{fig15} Dispersion of the `sub-peaks'
in the photoemission spectra of NiO (top) and CoO (bottom).
Also shown are the energies of the poles of the self-energy.}
\end{figure}
Lastly we discuss the fine structure of the TM3d derived bands.
This is shown in Figure \ref{fig15}.
The band structures of all three compounds have a similar structure:
at the top of the band structure there is a group of
dispersive bands which shows a rough similarity with the
upper group of bands in the LDA band structure (see Figure
\ref{fig1}), which have mainly TM 3d-character.
The total width of this band complex is reduced by a factor
of $\approx 0.5$ as compared to LDA. These bands have high spectral
weight and produce the intense peaks at the
top of the angle-integrated spectra for NiO and CoO.
Separated from this group of dispersive bands there is then a region
with many almost dispersionless bands with relatively
low spectral weight.  This overall structure can be understood
by considering the spectral representation of the
self energy (\ref{selfrep})
and the equation for the poles of the Green's function
\[
\omega + \mu - \epsilon_{\bm k} - Re\;\Sigma(\omega) = 0
\]
where we have considered the single band case for simplicity.
Since $ Re\;\Sigma(\omega)$ takes any value between
$\infty$ and $-\infty$ in between two successive poles $\zeta_\nu$
and $\zeta_{\nu+1}$ there is one band in between any two successive poles
of the self energy.
This implies that the distance between these two successive poles 
is an upper bound for the width of this band,
which may be viewed as a kind of correlation narrowing.
Moreover, if a pole $\zeta_\nu$ has only a small residuum
the resulting pole of the Green's function will be almost
`pinned' very close to it - as can be seen
repeatedly in Figure \ref{fig15}.
The topmost group of relatively strongly dispersive bands
then is actually above the topmost pole of $\Sigma(\omega)$
in the valence band region and the
dominant `gap opening peak' in the center of the
insulating gap, see Figure \ref{fig7}. 
Since the separation in energy between these peaks is large -
of the order of the insulating gap - these bands still have
an appreciable width.
The similarity with the LDA band structure is due to the
fact that the dispersion of these bands
is largely due to direct $d$-$d$-hopping,
which remains operative also when the self-energy is
included. While the fine structure of the valence band top
is not really resolved experimentally as yet, at least experiment
puts a quite low upper limit - $\le 0.5eV$ along $(100)$ -
on the width of the individual bands. The VCA would be consistent
with that. \\
The large number of dispersionless bands at more negative binding energy
is produced by the large number of densely spaced
poles of the self-energy. These are in turn the consequence of the
large number of CEF-split multiplet states in the TMO$_6$ cluster.
Interestingly, at least in the case of NiO where
detailed band mapping is available from ARPES, the experimental band
structure is quite consistent with this overall structure,
namely a group of dispersive bands at the top of the valence band
and essentially dispersionless bands at more negative binding energy.
The distance between the dispersive band complex and the
dispersionless bands is underestimated somewhat by VCA.\\
To conclude we compare the results of the VCA for NiO
with recent LDA+DMFT calculations\cite{Kunesetal_band,yinetal}. 
\begin{figure}
\includegraphics[width=\columnwidth]{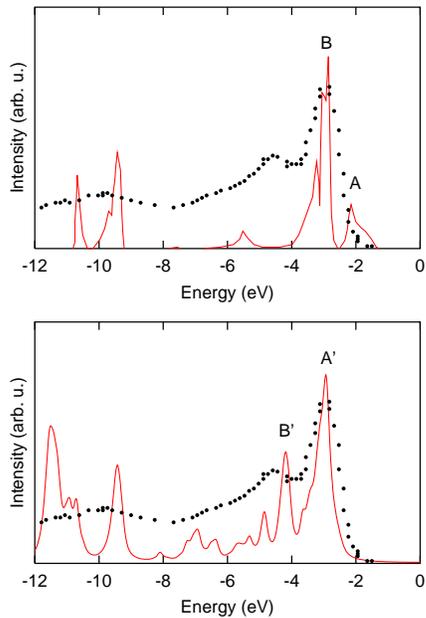}
\caption{\label{fig16} 
${\bm k}$-integrated $d$-like spectral weight obtained by
DMFT (top) (from Ref.\cite{yinetal}
and by VCA (bottom) compared to
XPS-data from NiO.}
\end{figure}
\begin{figure}
\includegraphics[width=\columnwidth]{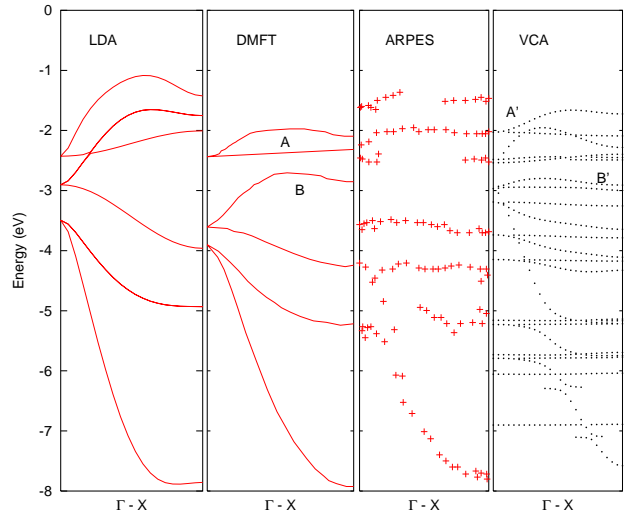}
\caption{\label{fig17} 
Bandstructure for NiO
along (100) as obatined from LDA, DMFT, ARPES and VCA.}
\end{figure}
Figure \ref{fig16} compares the ${\bf k}$-integrated
spectra, Figure \ref{fig17} shows the dipersion
along $(100)$. The DMFT-results are taken from 
Yin {\em et al.}\cite{yinetal} which have obatined essentially
identical results as Kunes {\em et al.}\cite{Kunesetal_band}.
While the ${\bf k}$-integrated spectra look similar
at first sight, comparison with the band structure shows that
there are major differences. In the DMFT spectrum the top
of the valence band is formed by a split off peak $A$
with low spectral
weight. This corresponds to the two topmost bands labelled $A$
in Figure \ref{fig17}. This form of the density of states is
actually reminiscent of the results of the three-body-scattering
theory of Manghi {\em et al.}\cite{Manghi}.
In the VCA spectrum the top of the valence band is formed by an 
intense peak $A'$, which
corresponds to the topmost band complex $A'$ in Figure \ref{fig17}.
The intense peak $B$ in the DMFT spectrum on the other hand
originates from the band $B$ in Figure \ref{fig17}.
The DMFT bands moreover show a rather obvious correspondence
with the LDA band structure, resulting in bands with quite
strong dispersion. As already noted
the VCA differs strongly from LDA
and shows a larger number of bands, with several of them being
practically dispersionless, i.e. corresponding
to localized electrons.\\
Comparing with experiment, the raw data
of Shen {\em et al.}\cite{Shen_long} show no indication for the split off
bands $A$ with low spectral weight as predicted by DMFT.
With the exception of the O2p derived
bands ARPES moreover shows no indication of the wide bands predicted by
DMFT - rather there is a number of dispersionless bands as expected
on the basis of the VCA. 
One may say that there are major differences between DMFT and VCA
so that further experiments might resolve this discrepancy.
\section{Conclusion}
To summarize, the variational cluster approximation due to Potthoff
allows to combine the powerful cluster configuration interaction
method for transition metal compounds due to Fujimori and Minami
with the field-theoretical work of Luttinger and Ward to implement a
variational scheme for the electronic self-energy and construct
an efficient band structure method for strongly correlated electron
compounds. As demonstrated above, a realistic band structure
and the full atomic multiplet interaction can be incorporated into
the Hamiltonian without problems, the system can be studied
at arbitrarily low temperatures and the Green's function be obtained
with arbitrary energy resolution. 
The key numerical problem of finding the stationary point of the
grand potential thereby can be solved efficiently
by a simple crossover procedure.
It has been shown that in the course of varying the self-energy
there may exist redundant degrees of freedom which leave the
grand potential almost unchanged. Such redundant degrees of freedom
can be eliminated by simply reducing the numer of parameters in the
reference system.\\
The results are quite encouraging in that there is good agreement
between the calculated  Green's function and electron spectroscopies
at least to the extent that ARPES data are available.
The good agreement also suggests that the band structure
of NiO, CoO and MnO is `Coulomb generated' in that the atomic multiplet
structure survives with minor modifications.
All in all the VCA appears to be a promising tool for the study of 
realistic models of correlated electron systems. The
possibility to treat the multiplet and CEF-splitting of the
various TM3d configurations more or less exactly
should make it possible to address magnetic or orbital ordering
phenomena in transition metal compounds.\\
I would like to thank M. Potthoff for many instructive
discussions.
\section{Appendix}
We show that $ln(det\;{\bf G}(\omega))$ is analytical
off the real axis, where ${\bf G}(\omega)$ can be either
the exact Greens function of the reference system or the
approximate Greens function form the VCA.
It is sufficient to proove that all eigenvalues
of ${\bf G}(\omega)$  have a nonvanishing imaginary part
for $\omega$ not on the real axis. This is prooved in turn
if we show that
\[
\langle {\bf v}| {\bf G}(\omega) |  {\bf v} \rangle =
\sum_{i,j} v^*_i\; G_{ij}(\omega)\; v_j
\]
has a nonvanishing imaginary part for any normalized ${\bf v}$.
For the exact Greens function we have - using the Lehman
representation -
\[
\langle {\bf v}| {\bf G}(\omega) |  {\bf v} \rangle =\frac{1}{Z}
\sum_{\nu',\nu} \frac{|C_{\nu'\nu}|^2}{\omega-(E_\nu-E_{\nu'})}
\left(e^{-\beta \epsilon_\nu } + e^{-\beta \epsilon_{\nu'}}\right)
\]
where $\epsilon_\nu = E_\nu - \mu N_\nu$ and
\[
C_{\nu'\nu} =\langle \nu' | \sum_{i} v_i c_{i} | \nu \rangle
\]
It follows that for $\omega$ in the upper (lower) half plane
all eigenvalues of ${\bf G}(\omega)$ have a negative (positive)
imaginary part and accordingly all eigenvalues of 
${\bf  G}^{-1}(\omega)$ have a positive (negative) imaginary part.
The imaginary part could only be zero if all $C_{\nu'\nu}$ were
zero which is unlikely to occur.
A similar proof has been given previously
by Dzyaloshinskii\cite{Dzyaloshinksii}.
Luttinger has shown that
the self-energy ${\bf \Sigma}(\omega)$ has a
spectral representation of the form
\[
{\bf \Sigma}(\omega) = {\bf g} +
\sum_\nu \frac{  {\bf S}_\nu}{\omega - \zeta_\nu}
\]
with a real ${\bf g}$\cite{Luttingerself}. Since ${\bf G}(\omega)$ is
Hermitean for real $\omega$ the matrices ${\bf S}_\nu$ are Hermitean too
moreover positive definite. Namely if one of the matrices ${\bf S}_\nu$ had
a negative eigenvalue $\lambda$ then 
${\bf  G}^{-1}(\zeta_\nu+i\epsilon)$ had the eigenvalue
$\frac{i}{\epsilon}\lambda$ plus terms which stay finite
as $\epsilon\rightarrow 0$, whereas we
have shown that all eigenvalues for $\omega$ in the upper half-plane
have positive imaginary parts. It follows immediately that
\[
\langle {\bf v} | \omega - {\bf \Sigma}(\omega) |  {\bf v} \rangle
\]
has a positive (negative) imaginary part for $\omega$ in the
upper (lower) half plane which prooves that all
eigenvalues of the aproximate Greens functions off the real
axis have nonvanishing imaginary parts as well.

\end{document}